# Design and Implementation of Butterworth, Chebyshev-I and Elliptic Filter for Speech Signal Analysis

Prajoy Podder
Department of ECE
Khulna University of
Engineering & Technology
Khulna-9203, Bangladesh

Md. Mehedi Hasan
Department of ECE
Khulna University of
Engineering & Technology
Khulna-9203, Bangladesh

Md.Rafiqul Islam
Department of EEE
Khulna University of
Engineering & Technology
Khulna-9203, Bangladesh

Mursalin Sayeed
Department of EEE
Khulna University of
Engineering & Technology
Khulna-9203, Bangladesh

## ABSTRACT
In the field of digital signal processing, the function of a filter is to remove unwanted parts of the signal such as random noise that is also undesirable. To remove noise from the speech signal transmission or to extract useful parts of the signal such as the components lying within a certain frequency range. Filters are broadly used in signal processing and communication systems in applications such as channel equalization, noise reduction, radar, audio processing, speech signal processing, video processing, biomedical signal processing that is noisy ECG, EEG, EMG signal filtering, electrical circuit analysis and analysis of economic and financial data. In this paper, three types of infinite impulse response filter i.e. Butterworth, Chebyshev type I and Elliptical filter have been discussed theoretically and experimentally. Butterworth, Chebyshev type I and elliptic low pass, high pass, band pass and band stop filter have been designed in this paper using MATLAB Software. The impulse responses, magnitude responses, phase responses of Butterworth, Chebyshev type I and Elliptical filter for filtering the speech signal have been observed in this paper. Analyzing the Speech signal, its sampling rate and spectrum response have also been found.

## Keywords
Impulse response, Magnitude response, Phase response, Butterworth filter, Chebyshev-I filter, Elliptical filter.

## 1. INTRODUCTION
Filters play an important role in the field of digital and analog signal processing and telecommunication systems. The traditional analog filter design consists of two major portions: the approximation problem and the synthesis problem. In early days the digital filter also faced accuracy problems because of the finite word length, but in the modern days because of the availability of 32 bit word lengths and floating point capabilities, digital filters are widely used. The primary functions of a filter are to confine a signal into a prescribed frequency band or channel or to model the input-output relation of a system such as a mobile communication channel, telephone line echo etc. [1]. Filters have many practical uses. To stabilize amplifiers by rolling off the gain at higher frequencies where extreme phase shifts may cause oscillations a single-pole low-pass filter (the integrator) is often used. Digital filters can be classified into two categories: FIR filter and IIR filter. Analog electronic filters consisted of resistors, capacitors and inductors are normally IIR filters [2]. On the other hand, discrete-time filters (usually digital filters) based on a tapped delay line that employs no feedback are essentially FIR filters. The capacitors (or inductors) in the analog filter have a "memory" and their internal state never completely relaxes following an impulse. But after an impulse response has reached the end of the tapped delay line, the system has no further memory of that impulse. As a result, it has returned to its initial state. Its impulse response beyond that point is exactly zero. IIR filter has certain properties such as width of the pass-band, stop-band, maximum allowable ripple at pass-band and maximum allowable ripple at stop-band [2], [3], [10]. A desired design of IIR filter can be done with the help of those properties [4]. The design of IIR digital filters with Butterworth, Elliptical filter responses, using MATLAB functions are based on the theories of bilinear transformation and analog filters [6]. So they are commonly used to approximate the piecewise constant magnitude characteristic of ideal HP, LP, BP and BS filters.

## 2. IIR FILTER
IIR filters can be usually implemented using structures having feedback (recursive structures). The present and the past input samples can be described by the following equation,

$$y(m) = \sum_{k=1}^{N} a_k y(m-k) + \sum_{k=0}^{M} b_k x(m-k) \ldots\ldots (2.1)$$

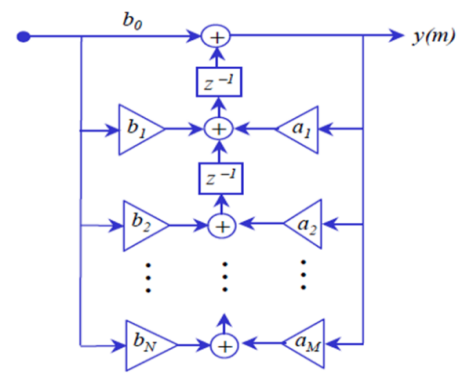

**Fig.1: An efficient realization of an IIR**

## 3. DIFFERENCE BETWEEN FIR & IIR

The main difference between IIR filters and FIR filters is that an IIR filter is more compact in that it can habitually achieve a prescribed frequency response with a smaller number of coefficients than an FIR filter. An IIR filter can become unstable, whereas an FIR filter is always stable [14] [15].

IIR filters have many advantages as follows [1]:-

i. Less number of arithmetic operations are required in IIR filter.

ii. There are shorter time delays in these filters.

iii. IIR Filters have similarities with the analog filters.

iv. Lesser number of side lobes in the stop band.

v. They are more susceptible to noises.

## 4. BUTTERWORTH FILTER

The Butterworth filter has a maximally flat response that is, no passband ripple and roll-off of minus 20db per pole. It is "flat maximally magnitude" filters at the frequency of $j\omega = 0$, as the first $2N - 1$ derivatives of the transfer function when $j\omega = 0$ are equal to zero [5]. The phase response of the Butterworth filter becomes more nonlinear with increasing N. This filter is completely defined mathematically by two parameters; they are cut off frequency and number of poles.

The magnitude squared response of low pass Butterworth filter is given by,

$$H(j\omega) = \frac{1}{1 + \left(\frac{\omega}{\omega_c}\right)^{2N}} \qquad \ldots (4.1)$$

Filter Selectivity,

$$F_s = \frac{N}{2\sqrt{2}\omega_c} \qquad \ldots (4.2)$$

Attenuation,

$$A = 10\log(1 + \left(\frac{\omega}{\omega_c}\right)^{2N}) \qquad \ldots (4.3)$$

The frequency response of the Butterworth filter is maximally flat in the passband and rolls off towards zero in the stopband [2]. When observed on a logarithmic bode plot the response slopes off linearly towards negative infinity. A first-order filter's response rolls off at −6 dB per octave (−20 dB per decade). A second-order filter's response rolls off at −12 dB per octave and a third-order at −18 dB. Butterworth filters have a monotonically varying magnitude function with ω, unlike other filter types that have non-monotonic ripple in the passband and the stopband.

Compared with a Chebyshev Type I filter or an Elliptic filter, the Butterworth filter has a slower roll-off and therefore will require a higher order to implement a particular stopband specification. Butterworth filters have a more linear phase response in the pass-band than Chebyshev Type I and Elliptic filters [11] [12] [13].

The Butterworth filter rolls off more slowly around the cut off frequency than the Chebyshev Type I and Elliptic filters without ripple. All of these filters are in fifth order.

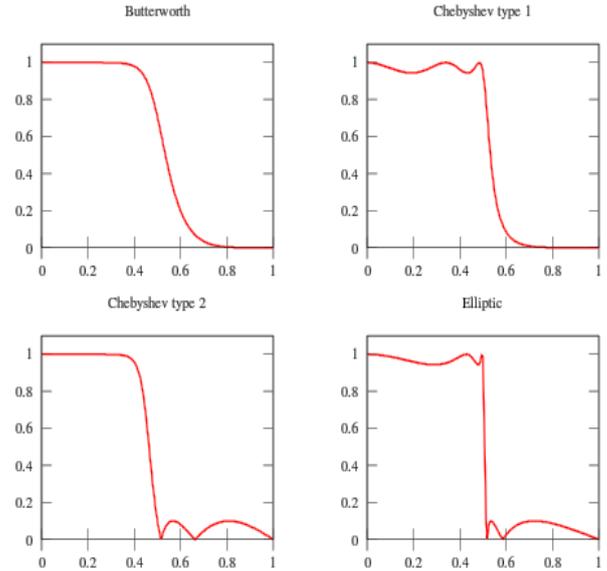

**Fig.2: Phase response of the various types of IIR filters**

## 5. CHEBYSHEV TYPE -I FILTER

The absolute difference between the ideal and actual frequency response over the entire passband is minimized by Chebyshev Type I filter by incorporating equal ripple in the passband. Stopband response is maximally flat. The transition from passband to stopband is more rapid than for the Butterworth filter. The magnitude squared Chebyshev type I response is:

$$H(j\omega) = \frac{1}{1 + \varepsilon^2 C_N^2(\frac{\omega}{\omega_p})} \qquad \ldots \ldots \ldots (5.1)$$

Where

$$C_N\left(\frac{\omega}{\omega_p}\right) = \begin{cases} \cos\left[N\cos^{-1}\left(\frac{\omega}{\omega_p}\right)\right], |\omega| \leq \omega_p \\ \cosh[N\cosh^{-1}\left(\frac{\omega}{\omega_p}\right)], |\omega| \geq \omega_p \end{cases} \qquad \ldots (5.2)$$

The magnitude squared response peaks occur in the pass band when $C_N\left(\frac{\omega}{\omega_p}\right) = 0$.

$$H(j\omega)_{min} = \frac{1}{1 + \varepsilon^2}$$

The ripple is often given in dB: Ripple = $20\log_{10}\sqrt{1 + \varepsilon^2}$

## 6. ELLIPTICAL FILTER

Elliptical filter can also be called as Cauer filters. Elliptic filters are equiripple in both the passband and stopband. They generally meet filter necessities with the lowest order of any supported filter type. For a given filter order, elliptic filters minimize transition width of the passband ripple and stopband ripple.

The magnitude response of a low pass elliptic filter as a function of angular frequency ω is given by [16],

$$H(j\omega) = \frac{1}{\sqrt{1 + \varepsilon^2 R_K^2\left(\xi, \frac{\omega}{\omega_p}\right)}} \qquad \ldots \ldots (6.1)$$

Where,

$R_k$ is the nth order elliptic rational function.

$\omega_0$ is the cut off frequency.

$\varepsilon$ is the ripple factor.

$\xi$ is the selectivity factor.

The value of the ripple factor specifies the passband ripple while the combination of the ripple factor and the selectivity factor specify the stopband ripple.

In the pass band, the elliptic rational function varies between zero and unity. The gain of the passband therefore will vary between 1 and $1/\sqrt{1+\varepsilon^2}$.

In the stopband, the elliptic rational function varies between infinity and the discrimination factor $L_k$.

$L_k = R_k(\xi,\xi)$

The gain of the stopband therefore will vary between 0 and $1/\sqrt{1+\varepsilon^2 L_k^2}$.

## 7. SIMULATION RESULTS

It is necessary to choose a suitable frequency range in order to design basic types of filters like low pass, High pass filters. Table 1 indicates the frequency specification for designing various types of IIR filter.

**Table-1: Frequency specification of filter design**

| Filter Type | Frequency (Hz) |
|---|---|
| Low pass | fp=2000Hz,fs=3000Hz |
| High pass | fp=3000Hz,fs=2000Hz |
| Band pass | $f_1$=1500,$f_2$=2000,$f_3$=1000,$f_4$=2500 |
| Band stop | $f_1$=1000,$f_2$=3000,$f_3$=1500,$f_4$=2500 |

Fig.3 shows the spectrum of the input signal i.e. the speech signal. The sampling rate of the speech signal is 8000 and the number of bits per sample is 16.

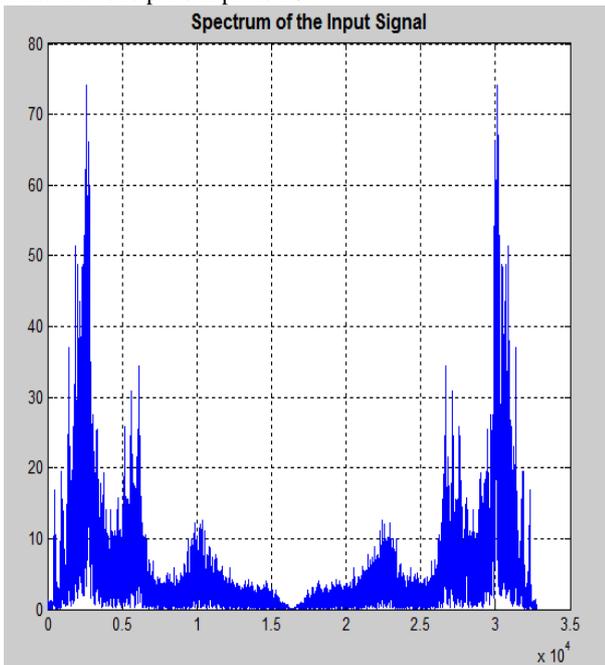

**Fig.3: Spectrum of the input signal**

Fig.4, 5, 6 and 7 shows the impulse response, magnitude response, phase response, pole zero plot and the output spectrum of the Butterworth low pass, high pass, band pass and band stop filter respectively. The pass band and stop band ripple are 2 and 35 respectively.

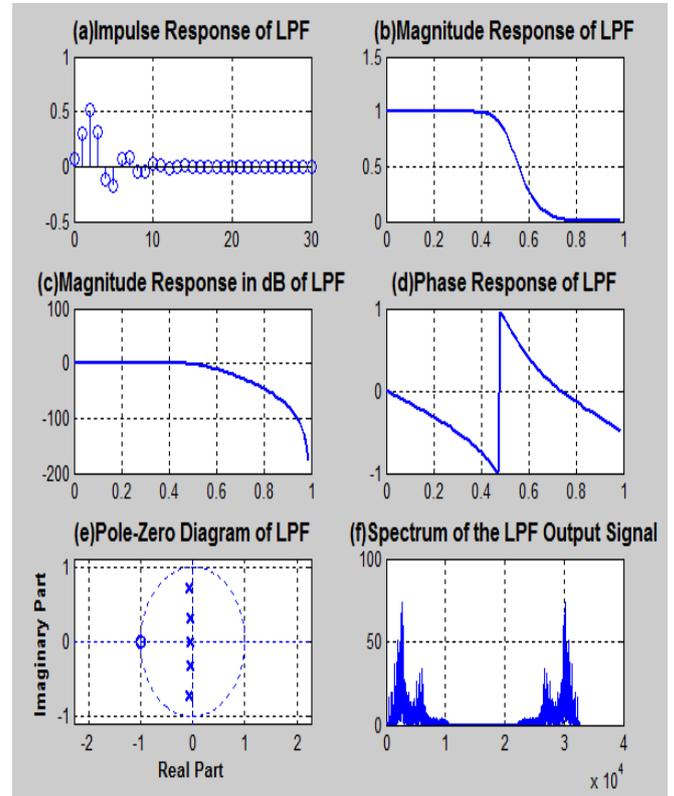

**Fig.4: Low pass Butterworth filter**

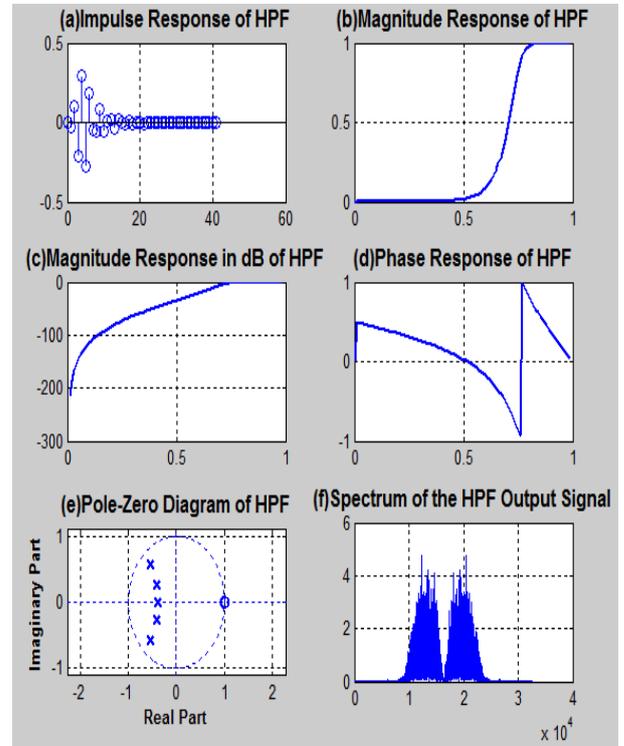

**Fig.5: High pass Butterworth filter**

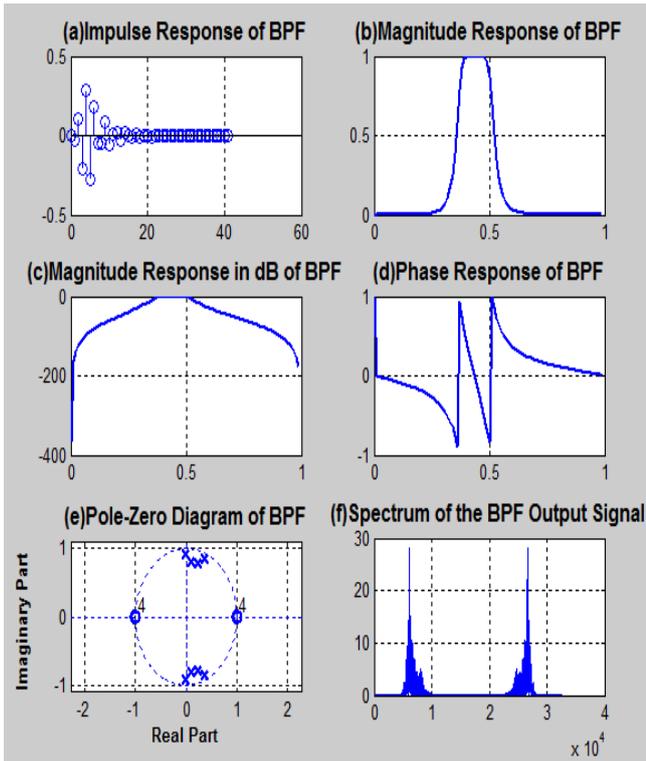

**Fig.6: Band pass Butterworth filter**

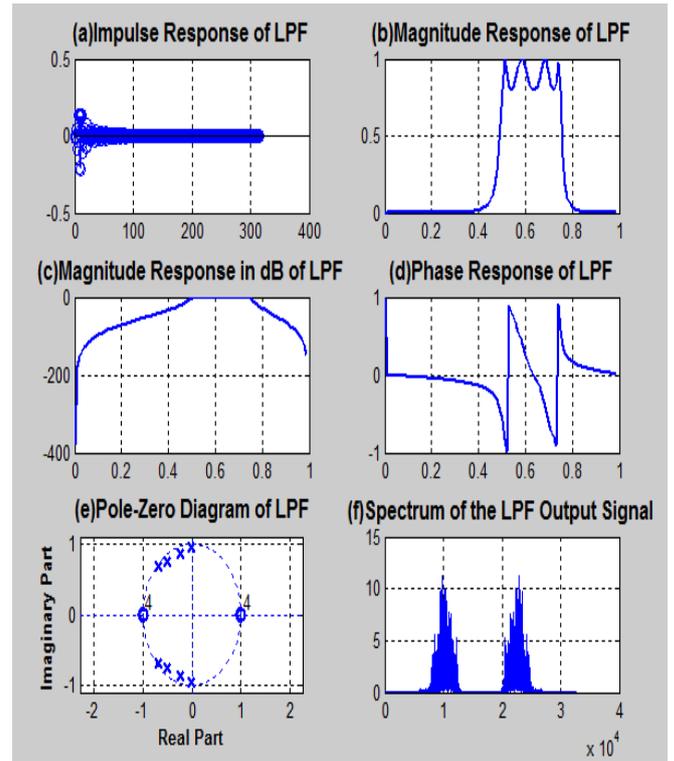

**Fig.8: Low pass hebyshev-1 filter**

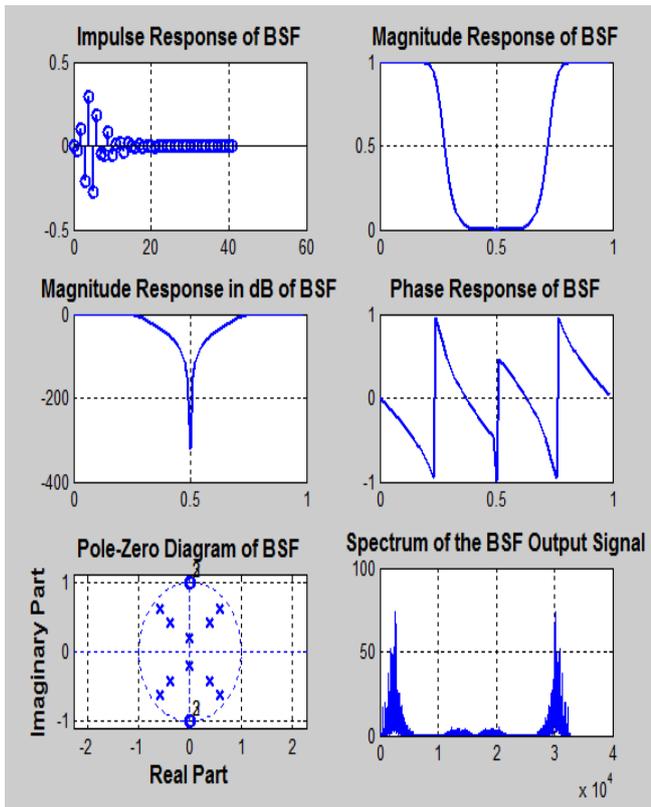

**Fig.7: Band stop Butterworth filter**

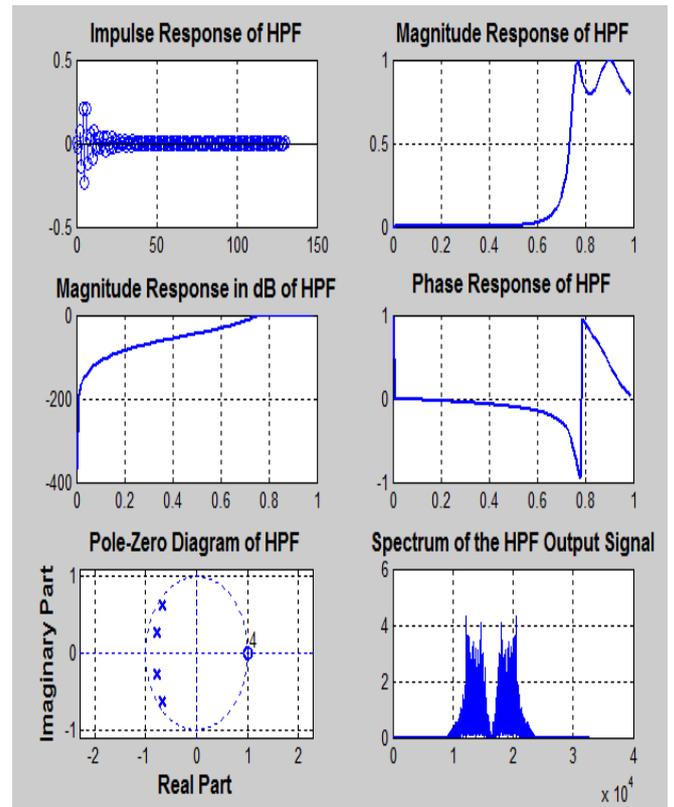

**Fig.9: High pass Chebyshev-1 filter**

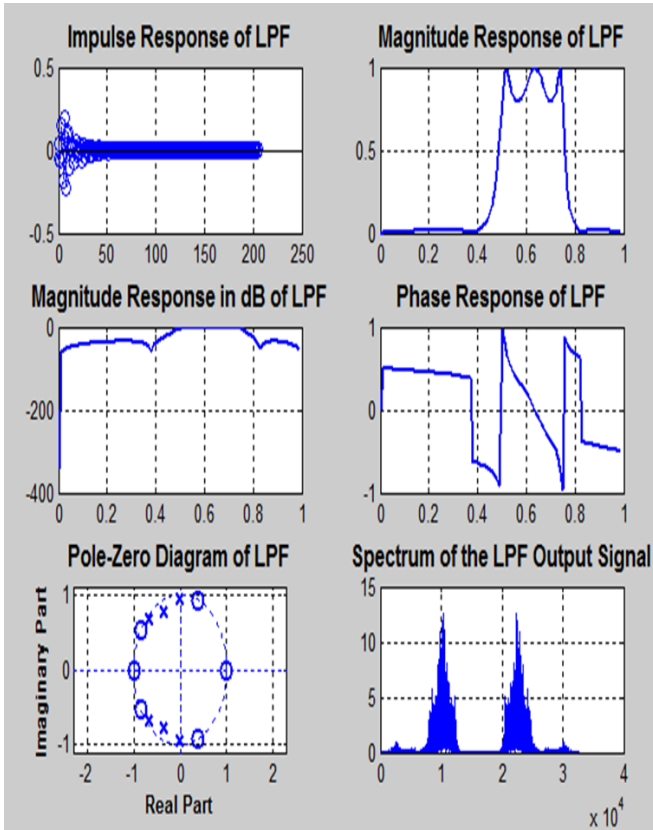

**Fig.10: Low pass Elliptical filter**

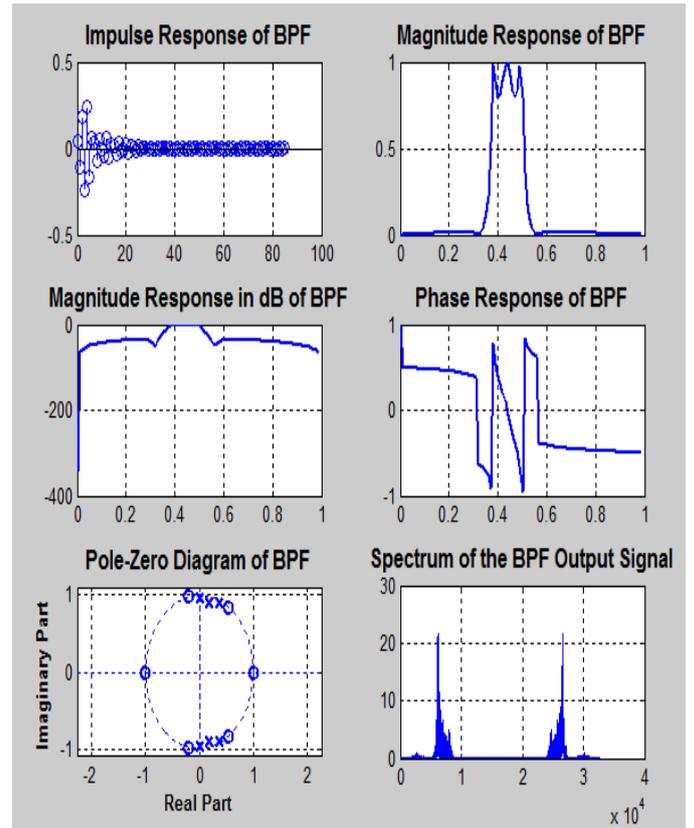

**Fig.12: Band pass elliptical filter**

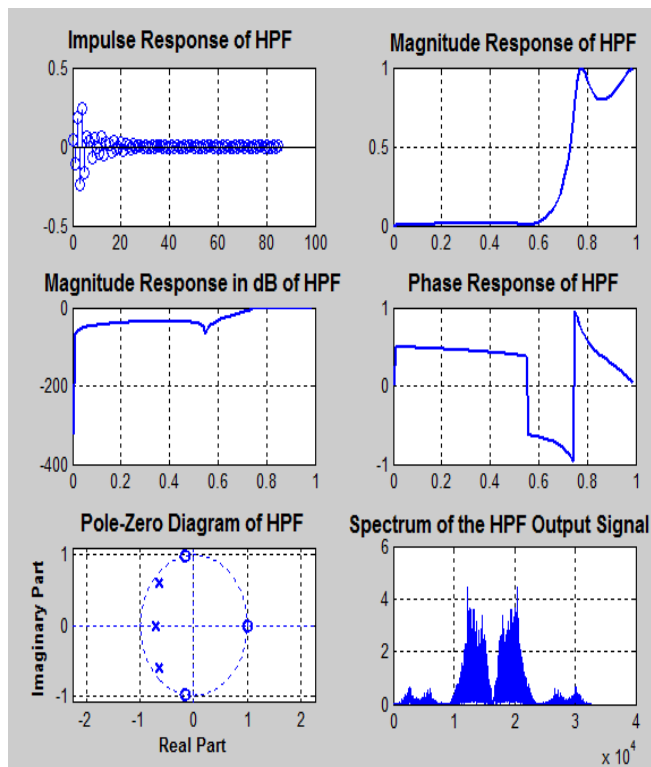

**Fig.11: High pass elliptical filter**

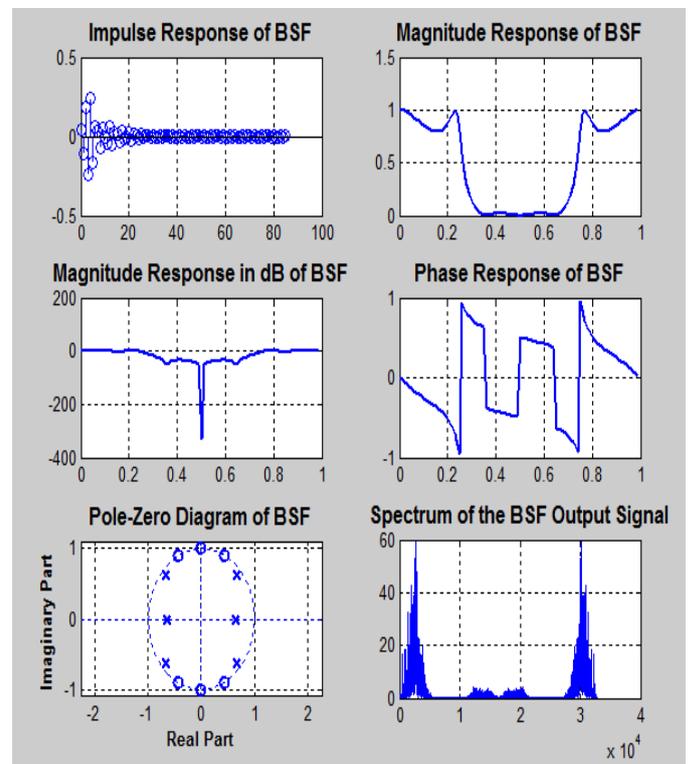

**Fig.13: Band stop elliptical filter**

Low Pass filter has pure poles. All other filters have complementary zeros .But, Butterworth LPF has complex poles. An important property of the Butterworth filter is the gain flatness in the passband. It has a realistically good phase response. Butterworth filter has a poor roll-off rate.

On the other hand Chebyshev has a better (steeper) roll-off rate because the ripple increases. Fig.8 and 9 indicates the impulse response, magnitude response, phase response, pole zero characteristics, output spectrum of the low pass and high pass Chebyshev type I filter respectively. Chebyshev filters have a poor phase response. The transfer function of a Chebyshev filter is characterized by a number of ripples in the passband .It has a smaller transition region than the same order Butterworth filter, at the expense of ripples in its pass band. Chebyshev type I filter minimizes the height of the maximum ripple. For the same filter order, the stopband attenuation is higher for the Chebyshev filter.

Compared to a Butterworth filter, a Chebyshev-I filter can achieve a sharper transition between the passband and the stopband with a lower order filter. The sharp transition between the passband and the stopband of a Chebyshev filter produces smaller absolute errors as well as faster execution speeds than a Butterworth filter. But it does not a good performance for speech signal analysis.

Fig 10, 11, 12 and 13 represents the basic characteristics of Elliptic LPF, HPF, BPF and BSF respectively. Its magnitude response in dB is normally negative.

The phase response is very non-linear. Elliptic filters offer steepest roll-off characteristics than Butterworth or Chebyshev I filter, but are equiripple in both the pass- and stopbands. Elliptic filter has a shorter transition region than the Chebyshev type I filter .The main reason is that it allows ripple in both the stop band and pass band. It is the addition of zeros in the stop band that causes ripples in the stop band.

Elliptical and Butterworth band stop filter have better performance for speech signal analysis i.e. voice signal can be smoothly listened.

## 8. CONCLUSION
Beside the mathematical comparison among Butterworth Chebyshev type I and elliptic filter, in this paper, these filters have also been encountered with the designed low pass, high pass, band pass and band stop infinite impulse response filter with a view to comparing their responses for different parameters like impulse response, magnitude response, phase response and pole zero characteristics all done using MATLAB simulation. The filter order, passband and stopband ripple are considered for the design of the IIR filter. It has been found that the Butterworth filter is the best compromise between attenuation and phase response. It has no ripple in the pass band or the stop band. The speech signals have also been encountered using MATLAB simulation, which was the special consideration, and compared the input and output spectrum of the signal.